\newcommand{\PRE}[1]{}
\documentclass
[aps,twocolumn,amsmath,preprintnumbers,amsfonts,superscriptaddress,showpacs]{revtex4}

\usepackage{bm}
\usepackage{epsfig}
\usepackage{graphicx}

\newcommand{\be}{\begin{equation}}
\newcommand{\ee}{\end{equation}}
\newcommand{\bea}{\begin{eqnarray}}
\newcommand{\eea}{\end{eqnarray}}
\newcommand{\nn}{\nonumber}
\newcommand{\ba}{\begin{array}}
\newcommand{\ea}{\end{array}}

\newcommand{\lsim}{
\mathrel{\hbox{\rlap{\hbox{\lower4pt\hbox{$\sim$}}}\hbox{$<$}}}}
\newcommand{\gsim}{
\mathrel{\hbox{\rlap{\hbox{\lower4pt\hbox{$\sim$}}}\hbox{$>$}}}}

\newcommand{\tev}{\text{TeV}}
\newcommand{\gev}{\text{GeV}}

\newcommand{\br}{{\tt Br}}
\newcommand{\fb}{\text{fb}}

\begin{document}

\preprint{MADPH-09-1537, UCRHEP-T471}

\title{
\PRE{\vspace*{1.5in}}
Six-Lepton $Z'$ Resonance at the Large Hadron Collider
\PRE{\vspace*{0.3in}} }
\author{Vernon Barger}
\affiliation{Department of Physics, University of Wisconsin, Madison, WI 53706, USA}
\author{Paul Langacker}
\affiliation{School of Natural Science, Institute for Advanced Study,
Einstein Drive, Princeton, NJ 08540, USA}
\author{Hye-Sung Lee}
\affiliation{Department of Physics and Astronomy, University of California, Riverside, CA 92521, USA}
\date{September, 2009}

\begin{abstract}
\PRE{\vspace*{.1in}} \noindent
New physics models admit the interesting possibility of a $Z'$ weak boson associated with an extra $U(1)$ gauge symmetry and a Higgs boson that is heavy enough to decay into a pair of $Z$ bosons.
Then $Z'$ production and decay via $Z' \to ZH \to ZZZ$ has a distinctive LHC signal that is nearly background free and reconstructs the $H$ and $Z'$ masses and widths.
The $Z'$ decay to 3 pairs of leptons is especially distinctive.
The $ZH$ decay mode exists even if the $Z'$ is decoupled from leptons, which motivates an independent 6-lepton resonance search regardless of the dilepton search results.
\end{abstract}


\maketitle

Many scenarios beyond the standard model (SM) have an extra Abelian $U(1)'$ gauge symmetry and an associated $Z'$ weak boson. (For a recent review, see Ref.~\cite{Langacker:2008yv}.)
If the $Z'$ has couplings to both quarks and leptons, it will be evident through a resonance in the Drell-Yan subprocess $q \bar q \to Z' \to \ell \bar \ell$ at hadron colliders \cite{Cvetic:1995zs,Carena:2004xs,Feldman:2006wb,Salvioni:2009mt}.
The current Tevatron lower bound on the mass of a $Z'$ that couples comparably to quarks and leptons is around $800$-$900 ~\gev$, with the exact bound depending on the model \cite{CDFdilepton,D0dilepton}.
There is a stringent bound from LEP on the $Z$-$Z'$ mixing angle: $|\alpha_{Z-Z'}| \lsim 10^{-3}$. (See Ref.~\cite{Erler:2009jh} and references therein.)
The LEP bound can be satisfied if the $Z'$ is sufficiently heavy ($M_{Z'}$ in the TeV range) or the property of the Higgs sector satisfies a certain condition\footnote{The exact condition depends on the details of the Higgs sector.
When there are two Higgs doublets, e.g., $\alpha_{Z-Z'} = \frac{1}{2} \arctan \left[ g_{Z} g_{Z'} \left( z[H_d] v_d^2 - z[H_u] v_u^2 \right)/(M_Z^2 - M_{Z'}^2) \right]$.
The LEP constraint can be satisfied for small $M_{Z'}$ when $v_u^2 / v_d^2 \sim z[H_d] / z[H_u]$ holds.}.

\begin{figure}[b]
\begin{center}
\includegraphics[width=0.27\textwidth]{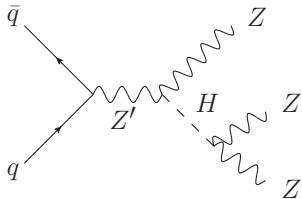}
\end{center}
\caption{6-lepton resonance without direct $Z'$ coupling to leptons.}
\label{fig:ZZZdiagram}
\end{figure}

At the LHC design energy of 14 TeV, the $Z' \to e^+e^-$ cross section for some representative models is of order $25 ~\fb$ for $M_{Z'}$ = 1.5 TeV and 1 fb for $M_{Z'} = 3 ~\tev$. (See, e.g., Table 2 of Ref.~\cite{Petriello:2008zr}.)
With an integrated luminosity of order 1-10 $\fb^{-1}$ in early LHC running, the clean $Z'$ dilepton Drell-Yan signal could be the first new physics discovery made at the LHC.
However, there exist models in which the $Z'$ does not have a direct coupling to leptons (e.g., Refs.~\cite{Babu:1996vt,Jung:2009jz}).
For such a leptophobic $Z'$ the Drell-Yan search is inapplicable, and masses considerably less than $800 ~\gev$ are possible.
A search for such a $Z'$ in the dijet channel will be challenging at the LHC \cite{CMS_dijet,ATL_dijet}.
Associated production of a $Z'$ with the SM $Z$ boson is an alternative search technique \cite{Petriello:2008pu}.
Our focus in this Letter is on a combined search at the LHC for the $Z'$ and a neutral $CP$-even Higgs boson ($H$) with mass $m_H > 2 M_Z$.

The physics of a TeV scale $U(1)'$ symmetry is intimately intertwined with the Higgs boson sector. (For example, see Ref.~\cite{Barger:2006dh} to see how the Higgs mass bound changes with $U(1)'$ in the supersymmetry framework.)
Precision electroweak constraints on the Higgs sector can also be relaxed because of mixing between $SU(2)$ doublet Higgs fields and the $SU(2)$ singlet fields ($S$) that are introduced to break the $U(1)'$ gauge symmetry \cite{Barger:2007im}.

We consider the $Z'$ decay to a $Z$ boson and a Higgs boson with subsequent decay $H \to ZZ$. (See Figure \ref{fig:ZZZdiagram}.)
It does not have to be the only or the lightest Higgs boson.
This mode exists even if the $Z'$ decouples from leptons.
The $Z' \to Z H$ channel has been studied in the literature \cite{Rizzo:1985kn,Nandi:1986rg,Baer:1987eb,Barger:1987xw,Gunion:1987jd,Deshpande:1988py,Hewett:1988xc}.
The $Z'$-$Z$-$H$ coupling is model dependent. 
To estimate its likely size, we consider the case of a single Higgs doublet.
Its kinetic term is
\bea
{\cal L}_\text{kin}
&=& \left|\left(\partial_\mu - \frac{i}{2} g_Z Z_\mu + i g_{Z'} z[H]
Z'_\mu\right) \frac{1}{\sqrt{2}} (H + v)\right|^2 \nn \\
&=& - g_Z g_{Z'} z[H] v H Z_\mu Z'^\mu + \cdots ,
\eea
where $g_Z$, $g_{Z'}$, $z[H]$, and $v$ are the coupling constants of the $Z$, and $Z'$, the $U(1)'$ charge of the Higgs doublet, and the Higgs vacuum expectation value, respectively.
The $Z'$-$Z$-$H$ coupling therefore depends on the $U(1)'$ charge of the Higgs doublet.
More generally, it also depends on mixing of states in the Higgs sector and on kinetic mixing.

In this Letter, we study the feasibility of the LHC to discover the $Z'$ as well as study the Higgs boson, using this vertex with the leptonic final states only without any missing energy.
We take an approach as model independent as possible, and show our results in two simple scenarios for the numerical illustration.

\begin{figure*}[tb]
\begin{center}
\includegraphics[width=0.329\textwidth]{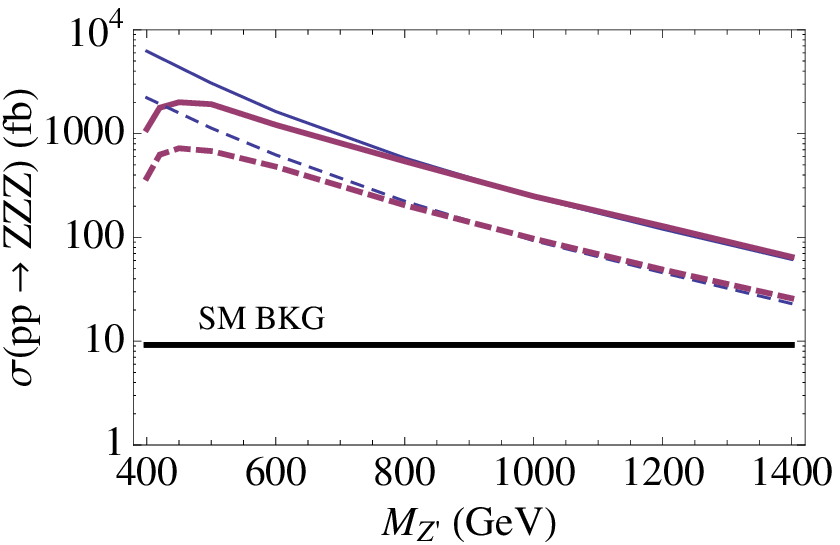}
\includegraphics[width=0.329\textwidth]{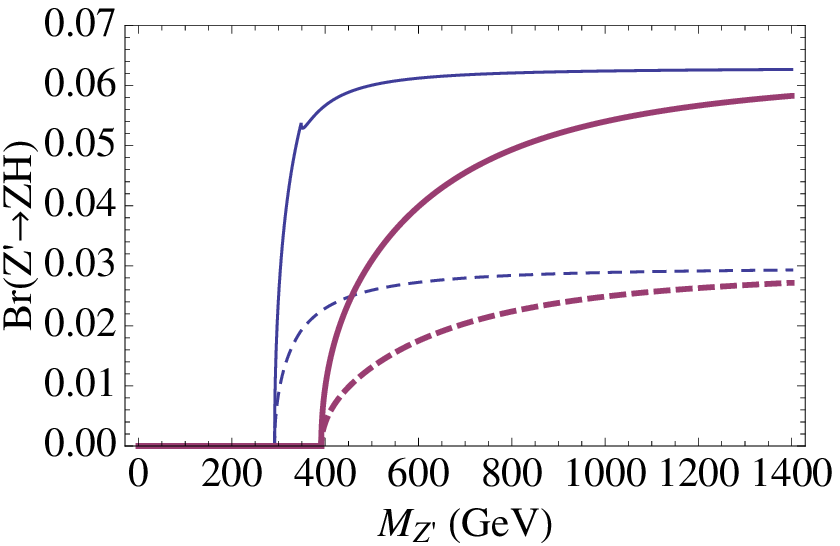}
\includegraphics[width=0.329\textwidth]{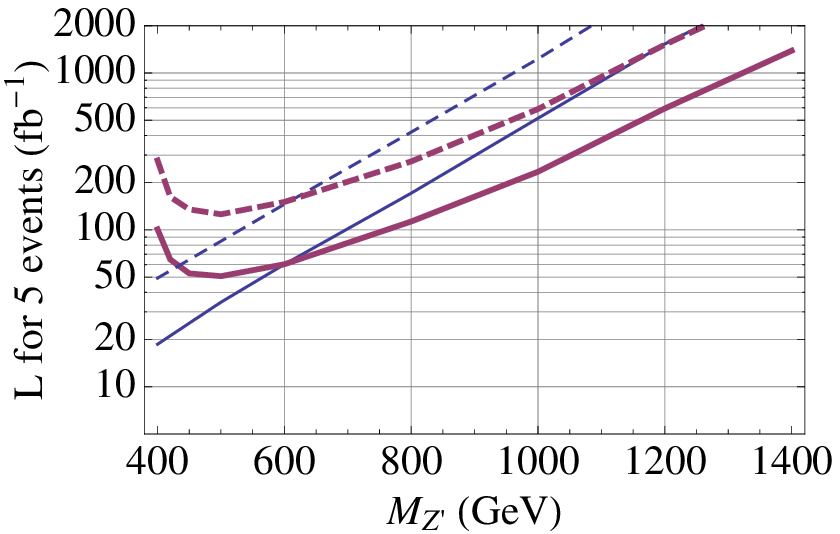}\\
(a)~~~~~~~~~~~~~~~~~~~~~~~~~~~~~~~~~~~~~~~~~~~~~~~~~~~~(b)~~~~~~~~~~~~~~~~~~~~~~~~~~~~~~~~~~~~~~~~~~~~~~~~~~~~(c)
\end{center}
\caption{
(a) $Z' \to ZH \to ZZZ$ cross section at the LHC (with $\sqrt{s} = 14 ~\tev$).
(b) $Z' \to Z H$ branching ratio.
(c) Required luminosity for five 6-lepton resonance events at the LHC (with $\sqrt{s} = 14 ~\tev$).
SM-like couplings (dashed) and leptophobic couplings (solid) are assumed with $m_H = 200 ~\gev$ (thin blue line) and $300 ~\gev$ (thick red line).
}
\label{fig:z2zh}
\end{figure*}

The decays of the Higgs boson are dependent on the Higgs mass and also on any mixing with singlets and other doublets.
The relevant $Z' \to Z H$ decay modes are $Z Z Z$, $Z W^\pm W^\mp$, which lead to the final states
\bea
Z Z Z&:& 6\ell,~ 4\ell + 2 j,~ 2\ell + 4 j,~ 6 j,~ 4\ell + \text{MET},~ \text{etc.} \nn \\
Z W W&:&  2\ell + 4 j,~ 6 j,~ 4\ell + \text{MET},~ \text{etc.} , \nn
\eea
where MET denotes missing transverse energy carried by neutrinos.

A dramatic $ZZZ$ signal is 3 pairs of opposite sign same flavor (OSSF) leptons that reconstruct three $Z$ bosons and then the Higgs boson and the $Z'$\footnote{For a discussion of $ZZZ$ production from an axial vector meson in a walking technicolor model, see Ref.~\cite{Belyaev:2008yj}.}.
Another particularly striking signal is a $b$-tagged pair and two pairs of leptons.

When the Higgs is lighter than $2 M_Z$, either $H$ or $Z$ can be off-shell; the event may be produced, but at a greatly reduced rate.

The SM background cross section from $t$-channel $q \bar q \to ZZZ$ is of order $10 ~\fb$ at $14 ~\tev$ \cite{Barger:1988sq}, which is small compared to the estimated $Z' \to Z H \to ZZZ$ signal. (See Figure \ref{fig:z2zh} (a).)
Thus the $6\ell$ decay mode is essentially free of backgrounds.
Since the $6\ell$ signal is very distinctive, only a few events would already confirm its existence.

The widths of both the $Z'$ and the Higgs boson can be determined from the invariant mass distributions of fully reconstructed events.

{\em 6-lepton resonance:}
Next we simulate the $Z' \to ZZZ$ signal at the LHC, implementing
typical experimental acceptance cuts and detector simulation.
Specifically, we consider the channel
\be
q \bar q \to Z' \to Z H \to 3 Z \to 6 \ell \label{eq:6leptons}
\ee with all intermediate decay particles on shell.
The 6-lepton resonance cross section is
\be
\sigma_{6\ell} \simeq \sigma(p p \to Z') \br(Z' \to Z H) \br(H \to Z Z) \br(Z \to \ell \bar\ell)^3 ,
\ee
where the major suppression comes from the triple $Z$ decay to light leptons, $\br(Z \to \ell \bar\ell) = \br(Z \to e^+ e^-) + \br(Z \to \mu^+ \mu^-) \simeq 0.067$.

For a single unmixed Higgs doublet\footnote{For models that include $SU(2)$ singlet scalars $S$ to break $U(1)'$ symmetry spontaneously, there can be mixing between the $H$ and $S$ through $H^\dagger H S^\dagger S$, which can modify the Higgs coupling even for the single Higgs doublet case.
However, we neglect this mixing effect in this Letter.}, the $ZH$ partial width is
\bea
&&\Gamma(Z' \to Z H) \\
&=&\frac{g_{Z'}^2 z[H]^2}{48 \pi} \lambda^{\frac{1}{2}}(1,x_Z,x_H) \left\{\lambda(1,x_Z,x_H) + 12 x_Z\right\} M_{Z'} , \nn
\eea
where $x_Z \equiv (M_Z/M_{Z'})^2$ and $x_H \equiv (m_H/M_{Z'})^2$, with $\lambda(x,y,z) = x^2 + y^2 + z^2 - 2xy - 2yz - 2zx$.
Its size can be comparable to or even larger than the $e^+e^-$ partial width
\be
\Gamma(Z' \to e^+ e^-) = \frac{g_{Z'}^2}{24 \pi} (z[e_L]^2 + z[e_R]^2) M_{Z'} .
\ee
Since the $ZH$ process does not require any direct coupling between the $Z'$ and leptons, it may be present even for a leptophobic $Z'$\footnote{This is to be compared with the 4-lepton $Z'$ resonance, where the $Z'$ decays to a sneutrino pair (SUSY search channel) \cite{Lee:2008cn}. It requires a leptonic coupling, and the dilepton $Z'$ resonance is unavoidable.}.
Therefore, a 6-lepton resonance for a leptophobic $Z'$ is possible.

For numerical illustrations, we use two examples of the $Z'$ couplings: (i) SM $Z$ couplings ({\em SM-like case}), (ii) leptophobic $Z'$ couplings ({\em leptophobic case}).
The SM-like case assumes $g_{Z'} = g_Z$.
The leptophobic case adopts $g_{Z'} = \sqrt{5/3} g \tan\theta_W \simeq 0.46$ and $z[u_L] = z[d_L] = 1/9$, $z[u_R] = -8/9$, $z[d_R] = 1/9$, $z[H_u] = -1$, $z[H_d] = 0$, while all leptons have vanishing $U(1)'$ charges\footnote{For the details of this model, which contains a $B_3$ residual discrete symmetry, see Refs.~\cite{Lee:2007fw,Lee:2007qx}. Though we chose zero charge for the leptons for the illustration, it is possible to assign small yet nonvanishing $U(1)'$ charges from a hypercharge shift.}.

Though the physical Higgs state depends on mixing among the Higgs doublets and singlets, for illustration we treat the model as effectively having a single Higgs doublet by taking no mixing and $v_u = v \simeq 246 ~\gev$\footnote{In the decoupling limit with $v_u \gg v_d$, there would not be much mixing between $H_u$ and $H_d$.}.
Multiple Higgses would lower the event rates and the dilution would depend on details of the Higgs sector.
Numerical results with our treatment will serve as the upper bound on the size of the signal.

\begin{figure}[tb]
\begin{center}
\includegraphics[width=0.23\textwidth]{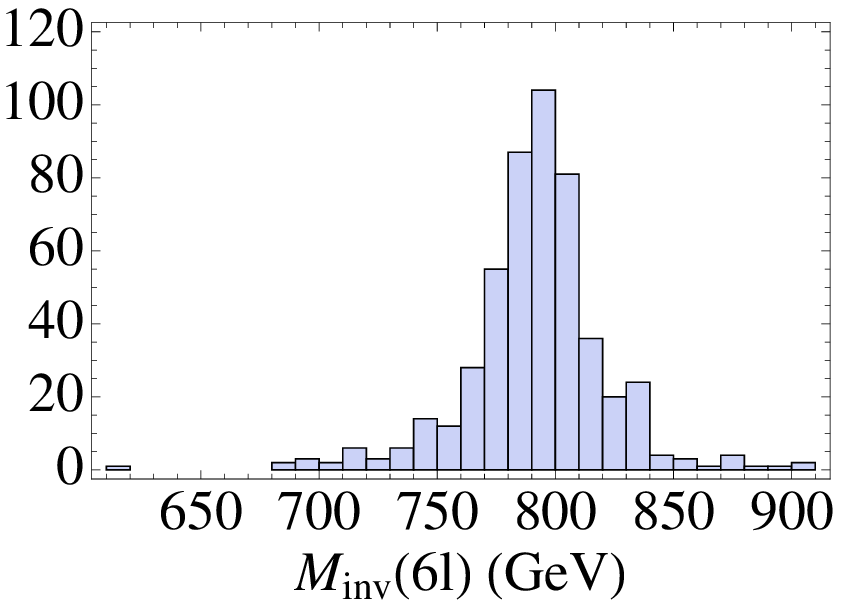} ~~
\includegraphics[width=0.23\textwidth]{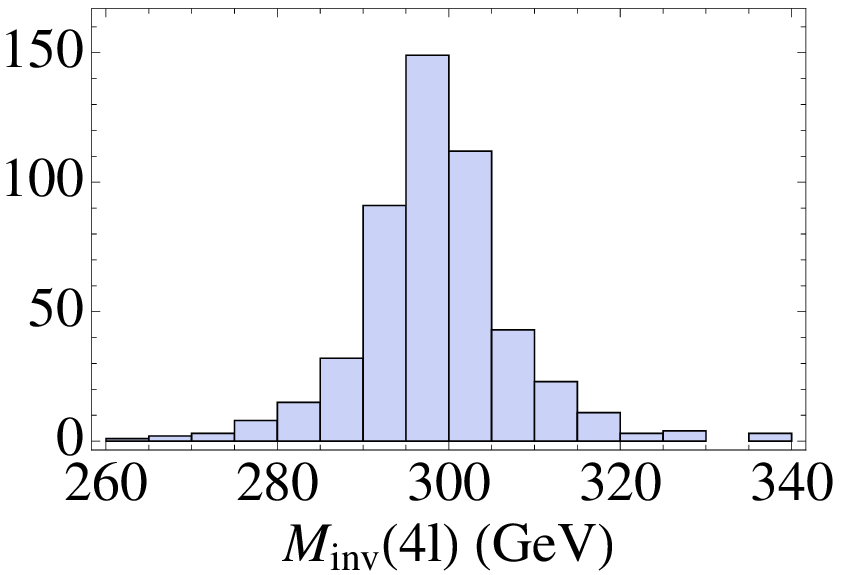} \\
(a) ~~~~~~~~~~~~~~~~~~~~~~~~~~~~~~~~~~~~~ (b) \\[1mm]
\includegraphics[width=0.30\textwidth]{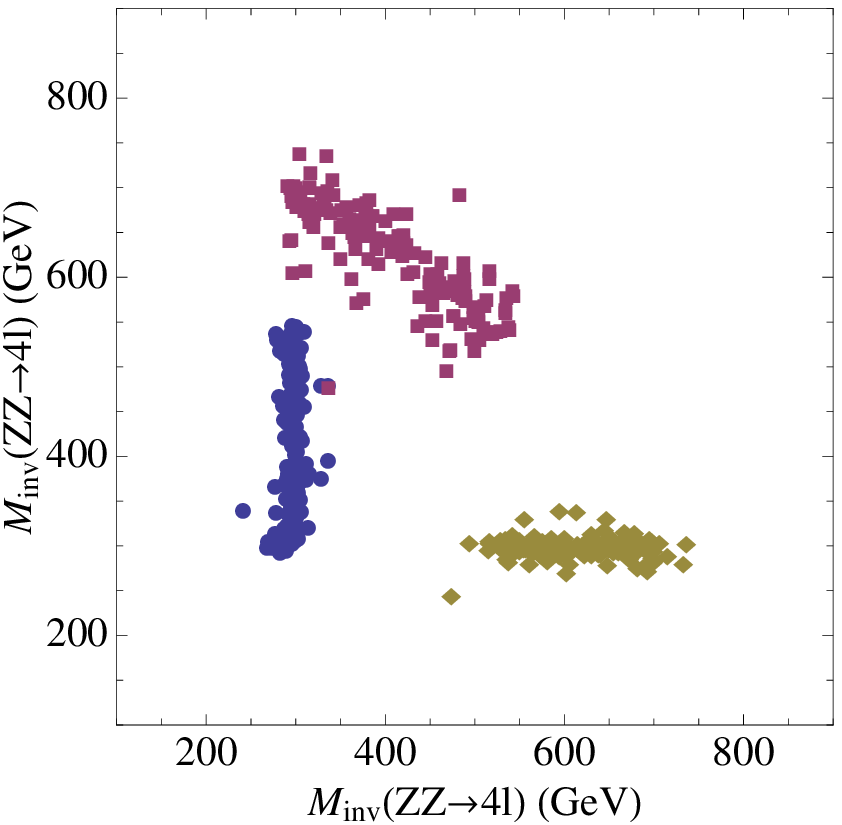} \\
(c)
\end{center}
\caption{
Reconstruction of the $Z'$ and Higgs boson for $M_{Z'} = 800 ~\gev$, $\Gamma_{Z'} = 16.6 ~\gev$, $m_H = 300 ~\gev$ and leptophobic couplings. 
(a) 6 leptons around the $Z'$ mass.
(b) 4 leptons around the Higgs mass.
(c) Dalitz plot revealing the Higgs mass.
The axes are 2 possible reconstructions for the $m_\text{inv} (4\ell)$.
There are 3 combinations of 2 pairs of $ZZ$.
These 3 combinations are sorted and plotted together in \{lightest, middle\} (blue circle), \{middle, heaviest\} (red box), and \{heaviest, lightest\} (yellow diamond).}
\label{fig:histogram}
\end{figure}

Figure \ref{fig:z2zh} (a) shows $\sigma(p p \to Z' \to ZZZ)$ at the LHC for $m_H = 200$ and $300 ~\gev$ in both models, and Figure \ref{fig:z2zh} (b) shows $\br(Z' \to Z H)$.
(There can also be a $Z' \to WW$ mode due to the $Z$-$Z'$ mixing.)
Larger $M_{Z'}$ would provide larger $\br(Z' \to Z H)$ for a given $m_H$, but the $Z'$ production cross section would be smaller.
Larger $m_H$ would decrease $\br(Z' \to Z H)$ but increase $\br(H \to Z Z)$.

The total decay widths are roughly $\Gamma_{Z'} \sim 0.03 M_{Z'}$ (SM-like case) and $\Gamma_{Z'} \sim 0.02 M_{Z'}$ (leptophobic case).
The leptophobic case has a larger $\br(Z' \to Z H)$ than the SM-like case because of the relatively large $z[H]$.

The $ZH$ decay requires a nonzero $U(1)'$ charge for the Higgs doublet.
A purely leptophobic model with a single Higgs with $z[H] \ne 0$ would not allow lepton Yukawa couplings.
However, this shortcoming can be overcome in multidoublet models (as in our leptophobic case example)\footnote{It is also possible to have suppressed couplings of the $Z'$ (KK mode of a neutral gauge boson) to the light leptons in warped extra dimension models, due to the small profile overlap.
(For example, see Ref.~\cite{Agashe:2007ki}.)}.

{\em Analysis:}
For numerical analysis, we use {\tt MadGraph} (Ver.~4.4.15), which adopts {\tt Pythia} and {\tt PGS} for hadronization and detector simulation\footnote{We use the detector parameters for CMS including (i) 70 cells for $\eta$ and $\phi$ in the calorimeter, (ii) a width 0.087 for the $\eta$ and $\phi$ calorimeter cells, (iii) $\Delta E / E = 0.01 + 0.03 \sqrt{E}$ for the electromagnetic calorimeter resolution.} \cite{Alwall:2007st}.
We use {\tt CTEQ6L} for parton distribution functions \cite{Pumplin:2002vw}, and take $M_{Z'}$ for the factorization scale.

$\br(Z' \to Z H)$ and $\br(H \to Z Z)$ depend on the details of the model, including the couplings of the Higgs, leptons, and possible exotic fields.
We illustrate using the two examples (SM-like case and leptophobic case).
For $\br(H \to Z Z)$, we use the SM value for a given $m_H$: $\br(H \to Z Z) \simeq 0.26$ (for $m_H = 200 ~\gev$), $0.31$ (for $m_H = 300 ~\gev$). 
In these examples, a $Z' \to 3Z$ branching fraction of order $1\%$ is achievable.

We impose the following cuts:
(i) pick the $6$ highest $p_T$ leptons ($\ell = e$, $\mu$),
(ii) require $p_T > 15 ~\gev$ and $|\eta| < 2.4$ for each $\ell$,
(iii) require the $6 \ell$ to be 3 pairs of light OSSF ($\ell \bar\ell
$), each of which satisfies $|m_\text{inv}(\ell \bar\ell) - M_Z| < 4
\, \Gamma_Z$.

Irreducible backgrounds to the 6-lepton resonance from $Z' \to ZZZ$ are negligibly small,
so we only require 5 events after the above cuts to claim discovery.

Figure \ref{fig:z2zh} (c) shows the luminosity needed for the discovery at the LHC with the design energy $\sqrt{s} = 14 ~\tev$.
The rather large required luminosity for $M_{Z'} \sim 400 ~\gev$ in the $m_H = 300 ~\gev$ case is due to the small $\br(Z' \to Z H)$, as Figure \ref{fig:z2zh} (b) shows.
In the leptophobic case with $m_H = 200$-$300 ~\gev$, the required luminosity for 5 events after the cuts is about $L \approx 60 ~\fb^{-1}$ for $M_{Z'} = 600 ~\gev$, and $L \approx 110$-$170 ~\fb^{-1}$ for $M_{Z'} = 800 ~\gev$.
The LHC design luminosity is of this order over $1$-$2$ years of running; also six-lepton events from the ATLAS and CMS experiments can be combined since the signal is background-free.

Therefore, it is possible to have such a signal at the LHC, which can (i) tell whether the Higgs doublet has a $U(1)'$ charge, (ii) detect a leptophobic $Z'$, and (iii) improve the study of the Higgs sector when combined with conventional Higgs search channels\footnote{For $m_H > 2 M_Z$, the SM-Higgs can be produced via gluon-gluon or vector boson fusion processes and decay to $ZZ$ and $WW$. It is discoverable at the LHC with $L \sim {\cal O}(10) ~\fb^{-1}$ unless unnaturally heavy. (See, e.g. Ref.~\cite{Ball:2007zza}.)}.
Adding other channels, such as $Z' \to ZZZ \to 4\ell + 2j$, would increase the reach.

Figure \ref{fig:histogram} (a) shows the reconstruction of the $Z'$ resonance from 6 leptons for $M_{Z'} = 800 ~\gev$ and $\Gamma_{Z'} = 16.6 ~\gev$ for an arbitrary number of events in the leptophobic case.

If the number of events is sufficiently large, a Dalitz plot can reveal the resonance in the 3-body decay.
We identify 3 OSSF dilepton pairs, each of which is from a $Z$ (at the risk of rare but possible wrong combinations).
Figure \ref{fig:histogram} (c) shows a relevant Dalitz plot\footnote{An additional Higgs with a different mass may alter the plots.}.
Each axis is the invariant mass of two $Z$'s.
The scatter plot shows a peak density at $m_\text{inv}(Z Z) = m_H$ from $H \to Z Z$, and reveals the Higgs mass at $m_H = 300 ~\gev$.

The Higgs boson can then be reconstructed, as in Figure \ref{fig:histogram} (b), by choosing 2 OSSF lepton pairs from the 6-lepton events whose invariant mass $m_\text{inv}(4\ell)$ is closest to the Higgs mass, while maintaining the condition of three $Z$ pole resonances.

If a 6-lepton resonance is found and the corresponding dilepton resonance is not found, the LEP constraint of $\alpha_{Z-Z'}$ would suggest that the Higgs structure may be more complicated than two Higgs doublets and singlets.
For example, in the two models illustrated above $\alpha_{Z-Z'}$ would be too large unless $M_{Z'} \gsim 2 ~\tev$.
This difficulty could be eliminated if there were a separate Higgs doublet dedicated for the lepton sector (e.g., Ref.~\cite{Ma:2001dn}).

{\em Summary and Outlook:}
A TeV scale $Z'$ is well motivated and closely related to the Higgs physics.
The 6-lepton resonance search should be performed independent of the dilepton $Z'$ search (even for the mass range excluded by the dilepton search).
A resonance of 6 light leptons ($e$, $\mu$) from $Z' \to ZZZ$ is effectively background-free, and is an excellent channel to search for a leptophobic $Z'$ without any ambiguity from the missing energy and jet misidentification.
Its discovery at the LHC would imply the existence of a Higgs boson with a significant $U(1)'$ charge in the mass range $2 M_Z \lsim m_H \lsim M_{Z'} - M_Z$.

{\em Acknowledgments:}
We are grateful to P. Konar, E. Ma, and M. Son for useful discussions.
We thank Aspen Center for Physics and BNL for hospitality.
This work was supported by DOE grants No. DE-FG02-95ER40896 (VB), No. DE-FG03-94ER40837 (HL), NSF grant No. PHY-0503584 (PL).




\begin{thebibliography}{99}
\bibitem{Langacker:2008yv}
P.~Langacker,
Rev.\ Mod.\ Phys. {\bf 81}, 1199 (2009).

\bibitem{Cvetic:1995zs}
  M.~Cvetic and S.~Godfrey,
  arXiv:hep-ph/9504216.
  
\bibitem{Carena:2004xs}
  M.~S.~Carena, A.~Daleo, B.~A.~Dobrescu and T.~M.~P.~Tait,
  Phys.\ Rev.\  D {\bf 70}, 093009 (2004).

\bibitem{Feldman:2006wb}
  D.~Feldman, Z.~Liu and P.~Nath,
  JHEP {\bf 0611}, 007 (2006).

\bibitem{Salvioni:2009mt}
  E.~Salvioni, G.~Villadoro and F.~Zwirner,
  JHEP {\bf 0911}, 068 (2009).

\bibitem{CDFdilepton}
T.~Aaltonen {\it et al.}  [CDF Collaboration],
Phys.\ Rev.\ Lett.\  {\bf 102}, 031801 (2009);
ibid., {\bf 102}, 091805 (2009).

\bibitem{D0dilepton}
[D0 Collaboration],
Note 5923-CONF (http://www-d0.fnal.gov).

\bibitem{Erler:2009jh}
  J.~Erler, P.~Langacker, S.~Munir and E.~R.~Pena,
  JHEP {\bf 0908}, 017 (2009).

\bibitem{Petriello:2008zr}
  F.~Petriello and S.~Quackenbush,
  Phys.\ Rev.\  D {\bf 77}, 115004 (2008).

\bibitem{Babu:1996vt}
K.~S.~Babu, C.~F.~Kolda and J.~March-Russell,
Phys.\ Rev.\  D {\bf 54}, 4635 (1996).

\bibitem{Jung:2009jz}
  S.~Jung, H.~Murayama, A.~Pierce and J.~D.~Wells,
  arXiv:0907.4112.
  
\bibitem{CMS_dijet}
K. Gumus, N. Akchurin, S. Esen and R.M. Harris, CMS Note 2006/070.

\bibitem{ATL_dijet}
S. Gonzalez de la Hoz, L. March and E. Roos, ATL-PHYS-PUB-2006-003.

\bibitem{Petriello:2008pu}
  F.~J.~Petriello, S.~Quackenbush and K.~M.~Zurek,
  Phys.\ Rev.\  D {\bf 77}, 115020 (2008).

\bibitem{Barger:2006dh}
V.~Barger, P.~Langacker, H.~S.~Lee and G.~Shaughnessy,
Phys.\ Rev.\  D {\bf 73}, 115010 (2006).

\bibitem{Barger:2007im}
  V.~Barger, P.~Langacker, M.~McCaskey, M.~J.~Ramsey-Musolf and G.~Shaughnessy,
  Phys.\ Rev.\  D {\bf 77}, 035005 (2008).

\bibitem{Rizzo:1985kn}
  T.~G.~Rizzo,
  Phys.\ Rev.\  D {\bf 34}, 1438 (1986).
  
\bibitem{Nandi:1986rg}
  S.~Nandi,
  Phys.\ Lett.\  B {\bf 181}, 375 (1986).

\bibitem{Baer:1987eb}
  H.~Baer, D.~Dicus, M.~Drees and X.~Tata,
  Phys.\ Rev.\  D {\bf 36}, 1363 (1987).

\bibitem{Barger:1987xw}
V.~D.~Barger and K.~Whisnant,
Phys.\ Rev.\  D {\bf 36}, 3429 (1987).

\bibitem{Gunion:1987jd}
J.~F.~Gunion, L.~Roszkowski and H.~E.~Haber,
Phys.\ Rev.\  D {\bf 38}, 105 (1988).
  
\bibitem{Deshpande:1988py}
N.~G.~Deshpande and J.~Trampetic,
Phys.\ Lett.\  B {\bf 206}, 665 (1988).

\bibitem{Hewett:1988xc}
  J.~L.~Hewett and T.~G.~Rizzo,
  Phys.\ Rept.\  {\bf 183}, 193 (1989).

\bibitem{Belyaev:2008yj}
  A.~Belyaev, R.~Foadi, M.~T.~Frandsen, M.~Jarvinen, F.~Sannino and A.~Pukhov,
  Phys.\ Rev.\  D {\bf 79}, 035006 (2009).

\bibitem{Barger:1988sq}
  V.~D.~Barger and T.~Han,
  Phys.\ Lett.\  B {\bf 212}, 117 (1988).

\bibitem{Lee:2008cn}
H.~S.~Lee,
Phys.\ Lett.\  B {\bf 674}, 87 (2009).
  
\bibitem{Lee:2007fw}
  H.~S.~Lee, K.~T.~Matchev and T.~T.~Wang,
  Phys.\ Rev.\  D {\bf 77}, 015016 (2008).

\bibitem{Lee:2007qx}
  H.~S.~Lee, C.~Luhn and K.~T.~Matchev,
  JHEP {\bf 0807}, 065 (2008).

\bibitem{Agashe:2007ki}
  K.~Agashe {\it et al.},
  Phys.\ Rev.\  D {\bf 76}, 115015 (2007).
    
\bibitem{Alwall:2007st}
J.~Alwall {\it et al.},
JHEP {\bf 0709}, 028 (2007).

\bibitem{Pumplin:2002vw}
J.~Pumplin, D.~R.~Stump, J.~Huston, H.~L.~Lai, P.~M.~Nadolsky and W.~K.~Tung,
JHEP {\bf 0207}, 012 (2002).

\bibitem{Ball:2007zza}
  G.~L.~Bayatian {\it et al.}  [CMS Collaboration],
  J.\ Phys.\ G {\bf 34}, 995 (2007).

\bibitem{Ma:2001dn}
  E.~Ma and G.~Rajasekaran,
  Phys.\ Rev.\  D {\bf 64}, 113012 (2001).

\end{thebibliography}
\end{document}